\documentclass[11pt,fleqn]{article}
\usepackage{appb2,epsfig}

\begin{document}
\null
\hfill WUE-ITP-99-007\\
\null
\hfill hep-ph/9904209\\
\vskip .8cm
\begin{center}
{\Large \bf Constraining the Sneutrino Mass in\\ 
Chargino Production and Decay
with\\ Polarized Beams
\\[.5em]
}
\vskip 2.5em

{\large  G. Moortgat-Pick,
H. Fraas}
\address{ 
Institut f\"ur Theoretische Physik, Am Hubland\\
D-97074 Universit\"at W\"urzburg, Germany}
\end{center}

\vspace*{1cm}
\begin{center}
{\bf Abstract}
\end{center}
Production and decay of gaugino-like charginos are crucially determined 
by sneutrino exchange. Therefore we
study the pair production of charginos 
$e^{+}e^{-}\to\tilde{\chi}^{+}_1\tilde{\chi}^{-}_1$ with polarized
beams and the subsequent decay 
$\tilde{\chi}^{-}_1\to\tilde{\chi}^0_1 e^{-} \bar{\nu}_e$, 
including the complete spin correlations between
production and decay. The spin correlations have strong influence on
the decay angular distribution and on the corresponding forward--backward 
asymmetry. We show for two representative scenarios for $\sqrt{s}=270$~GeV and
for $\sqrt{s}=500$~GeV that forward--backward asymmetries for
polarized beams are an important tool for constraining the 
sneutrino mass $m_{\tilde{\nu}_e}$.

\null
\setcounter{page}{0}
\clearpage

\newpage
\title{Constraining the Sneutrino Mass\\ 
in Chargino Production and Decay
with Polarized Beams\thanks{Presented by G. Moortgat-Pick at Cracow Epiphany
  Conference on {\it Electron Positron Colliders}, Cracow, Poland,
  January 5--10, 1999. Work supported by the German 
Federal Ministry for Research and Technology (BMBF) under contract number
05 7WZ91P (0).}
}

\author{G.~Moortgat--Pick \and H.~Fraas
\address{Institut f\"ur Theoret. Physik, Universit\"at W\"urzburg, Am
  Hubland,\\ D-97074 W\"urzburg}}
\maketitle

\vspace{0.5cm}
\begin{abstract}
Production and decay of gaugino-like charginos are crucially determined 
by sneutrino exchange. Therefore we
study the pair production of charginos 
$e^{+}e^{-}\to\tilde{\chi}^{+}_1\tilde{\chi}^{-}_1$ with polarized
beams and the subsequent decay 
$\tilde{\chi}^{-}_1\to\tilde{\chi}^0_1 e^{-} \bar{\nu}_e$, 
including the complete spin correlations between
production and decay. The spin correlations have strong influence on
the decay angular distribution and on the corresponding forward--backward 
asymmetry. We show for two representative scenarios for $\sqrt{s}=270$~GeV and
for $\sqrt{s}=500$~GeV that forward--backward asymmetries for
polarized beams are an important tool for constraining the 
sneutrino mass $m_{\tilde{\nu}_e}$.
\end{abstract}
  
\section{Introduction}
The search for supersymmetry (SUSY) is one of the main
goals of the present and future colliders. In particular an $e^{+}
e^{-}$ linear collider will be an excellent discovery machine for SUSY 
particles \cite{JLC}. Among them the charginos --
mixtures of W-inos and charged higgsinos -- are of particular
interest. They belong to the lightest SUSY particles and 
are produced with large cross sections if kinematically accessible. Their 
properties depend on the SU(2) gaugino mass  $M$, the higgsino mass
parameter $\mu$ and the ratio $\tan\beta=v_2/v_1$ of the
vacuum expectation values of the two neutral Higgs fields.

A number of studies addressed the determination of these fundamental
SUSY parameters from chargino production at $e^{+} e^{-}$ colliders
\cite{feng,etal,tsukamoto}. 
Recently a procedure was proposed to
determine the parameters $M$, $\mu$ and $\tan\beta$ 
independently from the decay dynamics \cite{choi}.
For this, however, one has to assume that the selectron sneutrino
mass $m_{\tilde{\nu}_e}$ is already known, e.g. from sneutrino pair 
production.

Since for gaugino-like charginos 
the cross section sensitively depends on the sneutrino
mass the possibility to measure $m_{\tilde{\nu}_e}$ in chargino pair 
production at an $e^{+}e^{-}$ collider has also been discussed 
\cite{feng,tsukamoto}.

In this paper we study the prospects for constraining
$m_{\tilde{\nu}_e}$ with suitably polarized beams 
by combining information from the cross
section 
\begin{equation}
\sigma_{e^{-}}=
\sigma(e^{+}e^{-}\to\tilde{\chi}^{+}_1 \tilde{\chi}^{-}_1)\times
BR(\tilde{\chi}^{-}_1\to\tilde{\chi}^0_1 e^{-}
\bar{\nu}_e)\label{sigma}
\end{equation} 
and the forward--backward asymmetry (FB--asymmetry)
\begin{equation}
A_{FB}=\frac{\sigma_{e^{-}}(\cos\Theta_{-}>0)-\sigma_{e^{-}}(\cos\Theta_{-}<0)}
{\sigma_{e^{-}}(\cos\Theta_{-}>0)+\sigma_{e^{-}}(\cos\Theta_{-}<0)}\label{afb}
\end{equation}
of the decay lepton $e^{-}$ from the decay
$\tilde{\chi}^{-}_1\to\tilde{\chi}^0_1 e^{-} \bar{\nu}_e$. In eq.(\ref{afb})
$\Theta_{-}$ is the angle between the incoming electron beam and the outgoing
$e^{-}$ in the laboratory system.
For constraining $m_{\tilde{\nu}}$ the simultaneous polarization of both
beams turns out to be very useful \cite{cracow98}.

In Section~2 the general formalism is presented and in Section~3
the numerical results are discussed.
\section{General formalism}
\vspace{-.3cm}
 The production process
$e^{+} e^{-} \to \tilde{\chi}^{+}_1 \tilde{\chi}^{-}_1$
contains contributions from $\gamma$, $Z^0$ and $\tilde{\nu}_e$ exchange
and the decay process, $\tilde{\chi}^{-}_1 \to
\tilde{\chi}^0_1 \ell^{-} \bar{\nu}_e$ 
($\tilde{\chi}^{+}_1 \to
\tilde{\chi}^0_1 \ell^{+} \nu_e$)
contains contributions from $W$, $\tilde{\ell}_L$  and
$\tilde{\nu}_e$ exchange. 
These processes have been studied in \cite{moor} properly taking into account
the spin correlations between chargino production and decay.

In the following the amplitude for the production (decay)
of $\tilde{\chi}^{-}_1$ with
helicity $\lambda_1$ is denoted by $P^{\lambda_1}$ ($D_{\lambda_1}$). 
All other helicity indices are suppressed.  
Then the amplitude of the combined process is
\begin{equation}
T=\Delta_{\tilde{\chi}^{-}_{1}} \sum_{\lambda_1} 
P^{\lambda_1} D_{\lambda_1}.\label{eq_1}
\end{equation}

In eq.~(\ref{eq_1}) 
$\Delta_{\tilde{\chi}^{-}_{1}}=1/[p^2_{\tilde{\chi}^{-}_{1}}
-m^2_{\tilde{\chi}^{-}_{1}}+i m_{\tilde{\chi}^{-}_{1}}
 \Gamma_{\tilde{\chi}^{-}_{1}}]^{-1}$, $p^2_{\tilde{\chi}^{-}_1}$,
$m_{\tilde{\chi}^{-}_{1}}$ and 
$\Gamma_{\tilde{\chi}^{-}_{1}}$  
denote the propagator, the
four--momentum squared, the mass and the width of
$\tilde{\chi}^{-}_1$. For this propagator we use the narrow width
approximation. 

The amplitude squared
$|T|^2=|\Delta_{\tilde{\chi}^{-}_{1}}|^2 \sum_{\lambda_1 \lambda^{'}_1}
\rho^{\lambda_1 \lambda_1'}_P
\rho^D_{\lambda_1' \lambda_1} \label{N}$
 is thus composed of the
unnormalized spin density matrix
 $\rho^{\lambda_1 \lambda_1'}_P=P^{\lambda_1} P^{\lambda_1' *}$
of $\tilde{\chi}^{-}_1$ and
 the decay matrix
 $\rho^{D}_{\lambda_1' \lambda_1}=D_{\lambda_1} D_{\lambda_1'}^{*}$. 
 Interference terms between
various helicity amplitudes preclude factorization into a production
factor  $\sum_{\lambda_1} |P^{\lambda_1}|^2$ times a decay factor
$\overline{\sum}_{\lambda_1} |D_{\lambda_1}|^2$. 
For the general case $e^{+}e^{-}\to \tilde{\chi}^{+}_i
\tilde{\chi}^{-}_j$ $(i, j=1, 2)$,
$\tilde{\chi}^{+}_i\to\tilde{\chi}^0_k \ell^{+} \nu_{\ell}$,
 $\tilde{\chi}^{-}_j\to\tilde{\chi}^0_l \ell^{-} \bar{\nu}_{\ell}$ $(k, 
l=1,\ldots,4)$
the complete analytical
formulae for polarized beams with full spin correlations between
production and decay are given in \cite{moor}.

\section{Numerical Results}
Our study is embedded in the Minimal Supersymmetric Standard
Model (MSSM) for two representative 
scenarios with  practically the same mass of the
lighter chargino and the lightest neutralino $\tilde{\chi}^0_1$,
respectively. The two scenarios differ,
however, significantly in the mixing character of the chargino.

The chargino mass eigenstates $\tilde{\chi}_i={\chi^+_i \choose
  \chi^{-}_i}$ are defined by $\chi^{+}_i=V_{i1} w^{+}+V_{i2} h^{+}$ 
and $\chi^{-}_i=U_{i1} w^{-}+U_{i2} h^{-}$. Here $w^{\pm}$ and
  $h^{\pm}$ are the two-component spinor fields of the W-ino and the charged
higgsinos, respectively. 
Furthermore $U_{ij}$ and $V_{ij}$ are the elements of the $2\times 2$ 
matrices which diagonalize the chargino mass matrix.

Below we give the W-ino and higgsino components of the two-component spinor
field $\chi^{-}_1$. Similarly we give the components of the neutralino
$\tilde{\chi}^0_1$ in the B-ino -- W-ino basis
$(\tilde{B}|\tilde{W}|\tilde{H}^0_a|\tilde{H}^0_b)$. For details see
\cite{haber,bartl86}. 
\setlength{\mathindent}{0cm}
\[ 
\mbox{\bf Scenario~A}:\quad M=152\mbox{~GeV}, M'=78.7\mbox{~GeV},
 \mu=316\mbox{~GeV}, \tan\beta=3.
\]
Here we have used the GUT relation $M'=\frac{5}{3} M \tan^2 \Theta_W$.
This set of parameters is inspired by the low $\tan\beta$ mSUGRA
reference scenario specified in \cite{blair}.
The eigenstates and the masses of the lighter chargino and of the LSP
$\tilde{\chi}^0_1$ are
\begin{eqnarray*}
\chi^{-}_1&=&(-0.91|+0.42)\quad\mbox{and}\quad
m_{\tilde{\chi}_1^{-}}=128\mbox{~GeV},\\
\tilde{\chi}^0_1&=&(-0.97|+0.15|-0.15|-0.15)\quad\mbox{and}\quad
m_{\tilde{\chi}^0_1}=71\mbox{~GeV}.\\
\end{eqnarray*}
\[
\mbox{\bf Scenario~B}:\quad M=400\mbox{~GeV}, M'=95\mbox{~GeV}, 
\mu=145\mbox{~GeV}, \tan\beta=3.
\]
Here the parameters are chosen thus that the chargino is higgsino-like
with only a small gaugino component. To obtain nearly the same masses as in
Scenario~A and an as 
large as possible higgsino component for $\tilde{\chi}^0_1$ 
we take $M$ and $M'$ as independent
parameters and do not use the GUT relation between them. 
The eigenstates and masses are
\begin{eqnarray*}
&&\chi^{-}_1=(-0.19|+0.98)\quad\mbox{and}\quad
m_{\tilde{\chi}_1^{-}}=129\mbox{~GeV},\\
&&\tilde{\chi}^0_1=(-0.82|+0.11|-0.45|-0.33)\quad\mbox{and}\quad
m_{\tilde{\chi}^0_1}=71\mbox{~GeV}.\\
\end{eqnarray*}
In both scenarios the LSP $\tilde{\chi}^0_1$ has a dominating B-ino component.

We vary the sneutrino mass $m_{\tilde{\nu}_e}$ between 80~GeV and
1000~GeV.
The mass of the left selectron $m_{\tilde{e}_L}$ is determined
by the relation 
\begin{equation}
\phantom{Scenario~A: \quad} m^2_{\tilde{e}_L}-m^2_{\tilde{\nu}_e}
=-M_W^2\cos 2\beta.
\end{equation} 
This sum rule is independent of any assumption about the gaugino mass
parameters and on the assumption of a universal scalar mass $m_0$ 
\cite{martin}.

The influence of spin correlations between production and decay on the
decay angular distributions is strongest near threshold.
Therefore numerical results are presented for energies near threshold
$\sqrt{s}=270$~GeV and for  
$\sqrt{s}=500$~GeV.

We have calculated for longitudinally polarized beams 
the cross section $\sigma_{e^{-}}$, eq.~(\ref{sigma}), and the FB--asymmetry 
$A_{FB}$, eq.~(\ref{afb}), of the decay electron.
For the polarization of the $e^{-}$ beam we assume $P_{-}=\pm85\%$
and for the polarization of the $e^{+}$ beam we take $P_{+}=\pm60\%$.
Simultaneous polarization of both beams is not only useful to enlarge
$\sigma_{e^{-}}$ but it can also have strong influence on $A_{FB}$ 
according to the mixing character of the charginos.
Since the sneutrino couples only to left-handed electrons the
polarization $P_{-}<0$ ($P_{+}>0$) for the $e^{-}$($e^{+}$) beam is
favourable for studying its properties.
\subsection{Cross sections}
For the gaugino-like scenario~A and $\sqrt{s}=270$~GeV the cross section 
$\sigma_{e^{-}}$ is shown in 
Fig.~1. 
For $m_{\tilde{\nu}_e}$ between 80~GeV and 400~GeV
it is very sensitive on $m_{\tilde{\nu}_e}$. The deep dip at 
$m_{\tilde{\nu}_e}\approx 130$~GeV 
is due to the destructive interferences between gauge boson and
slepton exchange \cite{bartl86,bartl92}. For 
$m_{\tilde{\nu}_e}<m_{\tilde{\chi}^{-}_1}$ the cross section 
increases owing to the two-body decay 
$\tilde{\chi}^{-}_1\to\tilde{\nu}_e e^{-}$. However, for
$\sqrt{s}=270$~GeV this effect is suppressed by destructive
interference effects. 

For higher energy $\sqrt{s}=500$~GeV, Fig.~2, the destructive
$Z^0$--$\tilde{\nu}_e$ interference 
in production is shifted to higher $m_{\tilde{\nu}_e}$ and 
the two-body decay threshold is more apparent.
For values of $m_{\tilde{\nu}_e}>200$~GeV
 the cross section $\sigma_{e^{-}}$ for
$\sqrt{s}=500$~GeV is smaller than for $\sqrt{s}=270$~GeV.

Beam polarization has strong influence on the magnitude of $\sigma_{e^{-}}$. 
For $P_{-}=-85\%$ and $P_{+}=+60\%$ it increases
by a factor of about 3, nearly independent of $\sqrt{s}$ and of 
$m_{\tilde{\nu}_e}$, Fig.~1 and Fig.~2.
For the opposite polarization configuration, $P_{-}=+85\%, P_{+}=-60\%$,
it would decrease by a factor of 0.06.  
 If only the electron beam is polarized 
$\sigma_{e^{-}}$ 
increases for $P_{-}=-85\%$ by a factor of 1.85 and for
$P_{-}=+85\%$ it decreases by a factor of 0.15.

In Fig.~3 and Fig.~4
we show $\sigma_{e^{-}}$ for the higgsino-like scenario~B. 
Since the dominating higgsino components do not couple to sneutrinos, 
the small gaugino admixture causes only a weak dependence on
$m_{\tilde{\nu}_e}$, Fig.~3. 
The destructive interference between $Z^0$ and $\tilde{\nu}_e$ exchange 
is suppressed by the dominating higgsino component.
For $m_{\tilde{\nu}_e}>m_{\tilde{\chi}^{-}_1}$ the cross section
is practically independent of
$m_{\tilde{\nu}_e}$. 
Only for $m_{\tilde{\nu}_e}<m_{\tilde{\chi}^{-}_1}$ it shows a strong
dependence on the sneutrino mass due to 
the two-body decay 
$\tilde{\chi}^{-}_1\to\tilde{\nu}_e e^{-}$.

For higher energy, $\sqrt{s}=500$~GeV, Fig.~4, 
$\sigma_{e^{-}}$ is smaller. The dependence on $m_{\tilde{\nu}_e}$, 
however, is essentially unchanged.  

Also for higgsino-like charginos
beam polarization has 
a strong influence on the magnitude of $\sigma_{e^{-}}$.
For $P_{-}=-85\%$ and $P_{+}=+60\%$  and for both values of $\sqrt{s}$,
Fig.~3 and Fig.~4, $\sigma_{e^{-}}$ increases
by a factor of about 2.5, which is somewhat smaller as for
the gaugino-like scenario~A.
For the opposite polarization, $P_{-}=+85\%, P_{+}=-60\%$, $\sigma_{e^{-}}$
decreases by a factor of 0.5.  
If only the $e^{-}$ beam is polarized with $P_{-}=-85\%$($P_{-}=+85\%$) 
the cross section  increases(decreases) 
by a factor of 1.6(0.4).
\subsection{Forward--backward asymmetry of the decay lepton}
In this section we study the FB--asymmetry $A_{FB}$
of the decay electron, eq.~(\ref{afb}),
for $m_{\tilde{\nu}_e}$ between 80~GeV and
1000~GeV. Since the angular distribution
of the chargino decay products is determined by the polarization of
the decaying chargino, the FB--asymmetry of the decay lepton  
strongly depends on the
spin correlations between production and decay.
The dependence of $A_{FB}$ on $m_{\tilde{\nu}_e}$, on the beam polarization
and on $\sqrt{s}$ is a result of the complex interplay
between production and decay.  

In scenario~A for $\sqrt{s}=270$~GeV the dependence of $A_{FB}$
on $m_{\tilde{\nu}_e}$ 
is shown in Fig.~5. For high $m_{\tilde{\nu}_e}>400$~GeV 
 it is about $8\%$ and is nearly independent of $m_{\tilde{\nu}_e}$.
For $m_{\tilde{\nu}_e}<400$~GeV, however, $A_{FB}$ shows
a complex and interesting mass dependence. 
The destructive $Z^0$--$\tilde{\nu}_e$ interference
in the production 
causes large $A_{FB}\approx 30\%$. Since this destructive
interference effect vanishes at about $m_{\tilde{\nu}_e}\approx 200$~GeV 
the FB--asymmetry decreases for higher values of $m_{\tilde{\nu}_e}$. 
The deep dip at $m_{\tilde{\nu}_e}\approx 140$~GeV is due to destructive
$W$--$\tilde{\nu}_e$ ($W$--$\tilde{e}$) interference in the decay process
$\tilde{\chi}^{-}_1\to \tilde{\chi}^0_1 e^{-} \bar{\nu}_e$. For
$m_{\tilde{\nu}_e}\le m_{\tilde{\chi}^{-}_1}$ the FB--asymmetry 
strongly increases up
to about $35\%$, owing to the two-body decay
$\tilde{\chi}^{-}_1\to\tilde{\nu}_e e^{-}$. 

For $\sqrt{s}=500$~GeV, Fig.~6,
large values of $A_{FB}$ up to $50\%$ can be reached in scenario~A.
The $Z^0$--$\tilde{\nu}_e$ interference effect in production
is shifted to higher $m_{\tilde{\nu}_e}$ so that just 
for $m_{\tilde{\nu}_e}\ge 250$~GeV $A_{FB}$ decreases 
from $48\%$ to $10\%$ for $m_{\tilde{\nu}_e}=1000$~GeV. 
For $\sqrt{s}=500$~GeV the $Z^0$--$\tilde{\nu}_e$ interference 
in the decay process is less important because 
for higher energies the characteristics of the production process 
is more and more dominating. 
Therefore at $m_{\tilde{\nu}_e}\approx 140$~GeV $A_{FB}$
decreases only to about $43\%$ in contrast to the deep dip at 
$m_{\tilde{\nu}_e}\approx 140$~GeV for $\sqrt{s}=270$~GeV. 
Again, for $m_{\tilde{\nu}_e}\le
m_{\tilde{\chi}^{-}_1}$ 
the two-body decay 
$\tilde{\chi}^{-}_1\to\tilde{\nu}_e e^{-}$ involves large 
$A_{FB}\approx 50\%$. 

Note that in the gaugino-like 
scenario~A $A_{FB}$ is nearly independent on beam polarization. 
This emerges as a consequence of the fact that
 in scenario~A the beam polarization 
changes the magnitude of $\sigma_{e^{-}}$ by a factor which is
nearly independent 
of $m_{\tilde{\nu}_e}$.  This factor cancels 
in $A_{FB}$, eq.~(\ref{afb}). 
Only for $P_{-}>0$ a small polarization effect can be observed, see
Fig.~5 and Fig.~6. 

In the higgsino-like scenario~B, Fig.~7, the FB--asymmetry for
$\sqrt{s}=270$~GeV 
is much smaller than in scenario~A, $A_{FB}\approx 5\%$
for $m_{\tilde{\nu}_e}\ge m_{\tilde{\chi}^{-}_1}$.
Only for $m_{\tilde{\nu}_e}\le m_{\tilde{\chi}^{-}_1}$ 
when the two-body decay
$\tilde{\chi}^{-}_1\to e^{-} \tilde{\nu}_e$ is open 
large $A_{FB}\approx 33\%$ can be reached for unpolarized
beams.

For $\sqrt{s}=500$~GeV $A_{FB}$ decreases,   
$A_{FB}\le 5\%$ for $m_{\tilde{\nu}_e}\ge m_{\tilde{\chi}^{-}_1}$ 
and $A_{FB}\approx 18\%$ for
$m_{\tilde{\nu}_e}\le m_{\tilde{\chi}^{-}_1}$, Fig.~8.

In this higgsino-like scenario~B the FB--asymmetry sensitively
depends on the beam polarization.
For $P_{-}=-85\%, P_{+}=+60\%$ 
the FB--asymmetry
increases for $m_{\tilde{\nu}_e}\le m_{\tilde{\chi}^{-}_1}$
up to $48\%$ 
for $\sqrt{s}=270$~GeV and up to $28\%$ for $\sqrt{s}=500$~GeV. 
If only the electron beam is polarized
with $P_{-}=+85\%$, 
one even obtains for $m_{\tilde{\nu}_e}\le m_{\tilde{\chi}^{-}_1}$
large negative FB--asymmetries, $A_{FB}\approx -20\%$($A_{FB}\approx -12\%$),
for $\sqrt{s}=270$~GeV($\sqrt{s}=500$~GeV).

\subsection{Constraining $m_{\tilde{\nu}_e}$}
In this section we study the prospects to constrain 
$m_{\tilde{\nu}_e}$ by measuring $\sigma_{e^{-}}$ and
$A_{FB}$ with different beam polarizations.

We assume that $\tan\beta$, $m_{\tilde{\chi}^{-}_1}$ and
$m_{\tilde{\chi}^0_1}$ are already known and we choose as an example the
gaugino-like scenario~A and the higgsino-like scenario~B. 

One obtains the largest cross sections for polarizations 
$P_{-}<0$,~$P_{+}>0$. We therefore assume that $\sigma_{e^{-}}$
has been measured for beam
polarizations $P_{-}=-85\%$, $P_{+}=+60\%$ and for energies
$\sqrt{s}=270$~GeV and $\sqrt{s}=500$~GeV with an error of 5\%.
We study also the correponding $A_{FB}$ for two combinations of
beam polarizations, $P_{-}=-85\%$, $P_{+}=+60\%$ and 
 $P_{-}=+85\%, P_{+}=0\%$. 
For the second case we use unpolarized positrons since also 
switching the $e^{+}$ polarization gives extremely small rates.
 \begin{center} 
{\it 3.3.1. Constraining $m_{\tilde{\nu}_e}$ for $\sqrt{s}=270$~GeV}
\end{center}
We assume that for $\sqrt{s}=270$~GeV and $P_{-}=-85\%$, 
$P_{+}=+60\%$ a cross section of
$\sigma_{e^{-}}=180 fb\pm 5\%$
has been measured.  
As can be seen from Fig.~1 for scenario~A 
and from Fig.~3 for scenario~B this cross section 
is compatible with two regions for $m_{\tilde{\nu}_e}$ 
in either case.

In scenario~A the cross section
is compatible with
small values of $m_{\tilde{\nu}_e}$ between 85~GeV and 89~GeV
and with high $m_{\tilde{\nu}_e}$ between 240~GeV and 255~GeV, 
Fig.~1. 
For the case of low $m_{\tilde{\nu}_e}$ the FB--asymmetry is between 
29\% and 32\% and for high $m_{\tilde{\nu}_e}$ 
between 19\% and 22\%, Fig.~5.
Accordingly $A_{FB}$ allows to discriminate between the two regions for 
$m_{\tilde{\nu}_e}$.
Changing the polarization to $P_{-}=+85\%$, $P_{+}=0\%$, the
FB--asymmetry gives no additional constraints for $m_{\tilde{\nu}_e}$, Fig.~5.

In scenario~B the cross section is 
compatible with small $m_{\tilde{\nu}_e}$ between
128~GeV and 140~GeV near the two-body decay threshold and with
high $m_{\tilde{\nu}_e}>250$~GeV, Fig.~3. 
For the case of low $m_{\tilde{\nu}_e}$ the FB--asymmetry is between
11\% and 29\% and for high $m_{\tilde{\nu}_e}$ 
$A_{FB}$ is less than $5\%$, Fig.~7. 
Again $A_{FB}$ allows to discriminate between the two regions for
$m_{\tilde{\nu}_e}$.

In our examples with $P_{-}=-85\%$, $P_{+}=+60\%$ 
the values of the FB--asymmetry for the low mass region of
scenario~B (11\%--29\%) overlaps with that for the low mass
region of scenario~A (29\%--32\%) as well as with that of the high mass 
region of scenario~A (19\%--22\%). 

However measuring $A_{FB}$ for a suitably
polarized electron beam and an unpolarized positron beam allows to
discriminate between scenario~A and scenario~B. 
For $P_{-}=+85\%$, $P_{+}=0\%$ the FB--asymmetry in the low mass region of
scenario~B is negative between $-5$\% and $-16\%$, whereas for scenario~A
the respective regions for $A_{FB}$ are unchanged, compare Fig.~5 and Fig.~7.
\begin{center}
{\it 3.3.2. Constraining $m_{\tilde{\nu}_e}$ for $\sqrt{s}=500$~GeV}
\end{center}
In the same manner we study 
$\sigma_{e^{-}}$ and $A_{FB}$ for $\sqrt{s}=500$~GeV.
We assume that
for polarized beams with $P_{-}=-85\%$, $P_{+}=+60\%$ the cross section
$\sigma_{e^{-}}=130 fb\pm 5\%$ has been measured.  
As can be seen from Fig.~2 for scenario~A and Fig.~4 
for scenario~B this cross section 
is again compatible with two regions for $m_{\tilde{\nu}_e}$ in 
either case.

In scenario~A $\sigma_{e^{-}}$
is compatible with small 
values of $m_{\tilde{\nu}_e}$ between $161$~GeV and $175$~GeV and 
with high values of $m_{\tilde{\nu}_e}$ between $420$~GeV and $450$~GeV, 
Fig.~2. 
For small $m_{\tilde{\nu}_e}$ the 
FB--asymmetry is about $45\%$ whereas 
for high $m_{\tilde{\nu}_e}$ 
it is between $32\%$ and $35\%$, clearly 
distinguishable from the case of
low $m_{\tilde{\nu}_e}$, Fig.~6. As
compared to the case of lower
energies, $\sqrt{s}=270$~GeV, the values of 
$A_{FB}$ are higher and restricted to 
smaller regions.
Changing the beam polarization to $P_{-}=+85\%$, $P_{+}=0\%$ 
the FB--asymmetry gives practically
no additional constraints for $m_{\tilde{\nu}_e}$, Fig.~6. 

In scenario~B $\sigma_{e^{-}}$ is compatible with
small $m_{\tilde{\nu}_e}$ in the region between $122$~GeV and $128$~GeV 
near the two-body decay threshold and with 
high $m_{\tilde{\nu}_e}>750$~GeV, Fig.~4. 
For low $m_{\tilde{\nu}_e}$ the FB--asymmetry
is between $10\%$ and $28\%$ whereas for high values of $m_{\tilde{\nu}_e}$ 
it is rather small, $A_{FB}<3\%$. 
These two regions for $m_{\tilde{\nu}_e}$ 
can clearly be distinguished by 
measuring $A_{FB}$, Fig.~8.

Similar as for lower energies, $\sqrt{s}=270$~GeV, the regions for 
$A_{FB}$ for the case of high $m_{\tilde{\nu}_e}$ in scenario~A and for the
case of $m_{\tilde{\nu}_e}\approx m_{\tilde{\chi}^{-}_1}$ in scenario~B 
are not well separated. 
Again  a suitably polarized $e^{-}$ beam and an unpolarized $e^{+}$
beam allow to discriminate between scenario~A and scenario~B.
For $P_{-}=+85\%$, $P_{+}=0\%$ the FB--asymmetry is negative between
$A_{FB}=-8\%$ and $A_{FB}=-12\%$ in scenario~B, whereas it has large
positive values between $A_{FB}=31\%$ and $A_{FB}=34\%$ in scenario~A, 
compare Fig.~6 and Fig.~8.
\section{Summary and Conclusions}
For the determination of the MSSM parameters $M$, $\mu$ and
$\tan\beta$ from chargino production, the sneutrino mass plays a
crucial role. With regard to the determination of the sneutrino mass we
have studied chargino pair production
$e^{+}e^{-}\to\tilde{\chi}^{+}_1\tilde{\chi}^{-}_1$ with polarized
beams and the subsequent leptonic decay
$\tilde{\chi}^{-}_1\to\tilde{\chi}^0_1 e^{-} \bar{\nu}_e$ taking into
account the complete spin correlations between production and decay.
The spin correlations are crucial for the decay angular
distributions and on the forward--backward asymmetries of
the decay leptons.

We have presented a method for constraining $m_{\tilde{\nu}_e}$
if  $\tan\beta$ and the masses of $\tilde{\chi}^{-}_1$ and
of $\tilde{\chi}^0_1$ are known.  
Accordingly we choose two representative scenarios with a dominating
gaugino and with a dominating higgsino component of
$\tilde{\chi}^{-}_1$, respectively.
We have studied the prospect to constrain 
the sneutrino mass by measuring polarized cross sections and
FB--asymmetries of the decay leptons in the laboratory system. 

Simultaneous polarization of both beams 
is not only important for enlarging the cross section but it
can also have rather strong 
 influence on the FB--asymmetry of the decay
electron depending on 
the mixing character of the charginos. 
For charginos with dominating gaugino character the FB--asymmetry 
is practically inpendent of the beam polarization.
But for higgsino-like charginos 
beam polarization can considerably influence the FB--asymmetry.

In our examples $\sigma_{e^{-}}$ measured for
the beam polarization $P_{-}=-85\%$, $P_{+}=+60\%$ with
an error of 5\%, e.g., is 
compatible with two regions for $m_{\tilde{\nu}_e}$. 
For gaugino-like charginos as well as for higgsino-like ones
the corresponding FB--asymmetry allows to
distinguish between the two regions of $m_{\tilde{\nu}_e}$. 
Beyond that the two mixing scenarios can be distinguished by
additional measurement of the FB--asymmetry with a right polarized
$e^{-}$ beam. Then the sign of $A_{FB}$ changes 
for higgsino-like charginos. 

We have presented numerical results for $\sqrt{s}=270$~GeV, near the
chargino production threshold, and for $\sqrt{s}=500$~GeV. 
For gaugino-like charginos
$m_{\tilde{\nu}_e}$ can be constrained up to a few GeV, for
higgsino-like charginos this is possible if $m_{\tilde{\nu}_e}$ is in
the vicinity of the chargino mass.
Obviously a complete MC study with inclusion of experimental cuts 
would be indispensable for realistic predictions.

\vspace{.4cm}
G.~M.--P.\ thanks M.~Jezabek and the other organizers of the Epiphany
Conference for the extremely hostly and 
friendly atmosphere during the Conference.
We are grateful to V.~Latussek for his support in the development of
the numerical program. G.M.--P.\ was supported by {\it Stiftung f\"ur
  Deutsch--Polnische Zusammenarbeit} and by {\it Friedrich--Ebert--Stiftung}.


\begin{picture}(10,8)
\put(4,0){\includegraphics{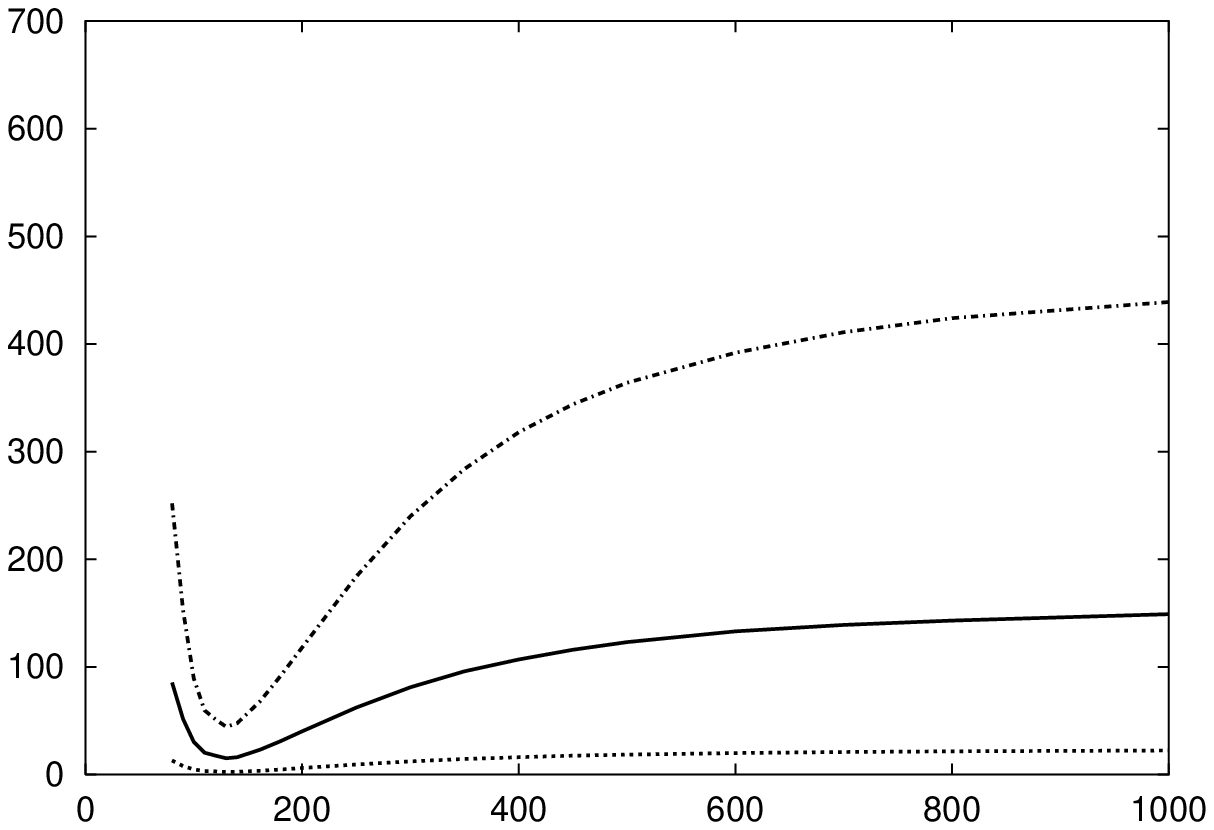}}
\put(290,-175){$ \scriptstyle m_{\tilde{\nu}_e}/GeV$}
\put(0,-18){$ \scriptstyle \sigma_{e^{-}} /fb$}
\put(0,-220)
{\parbox{11.5cm}{Fig.~1: Cross section $\sigma_{e^{-}}$ (see
    eq.~(\ref{sigma})) 
    for $\sqrt{s}=270$~GeV in scenario~A for unpolarized beams (solid
    line), with  only $e^{-}$ beam polarized $P_{-}=+85\%$ (dotted
    line) and with both beams polarized 
    $P_{-}=-85\%$, $P_{+}=+60\%$ (dash-dotted line).}\label{sit270_A}}
\end{picture}

\vspace{8.5cm}
\begin{picture}(10,8)
\put(4,0){\includegraphics{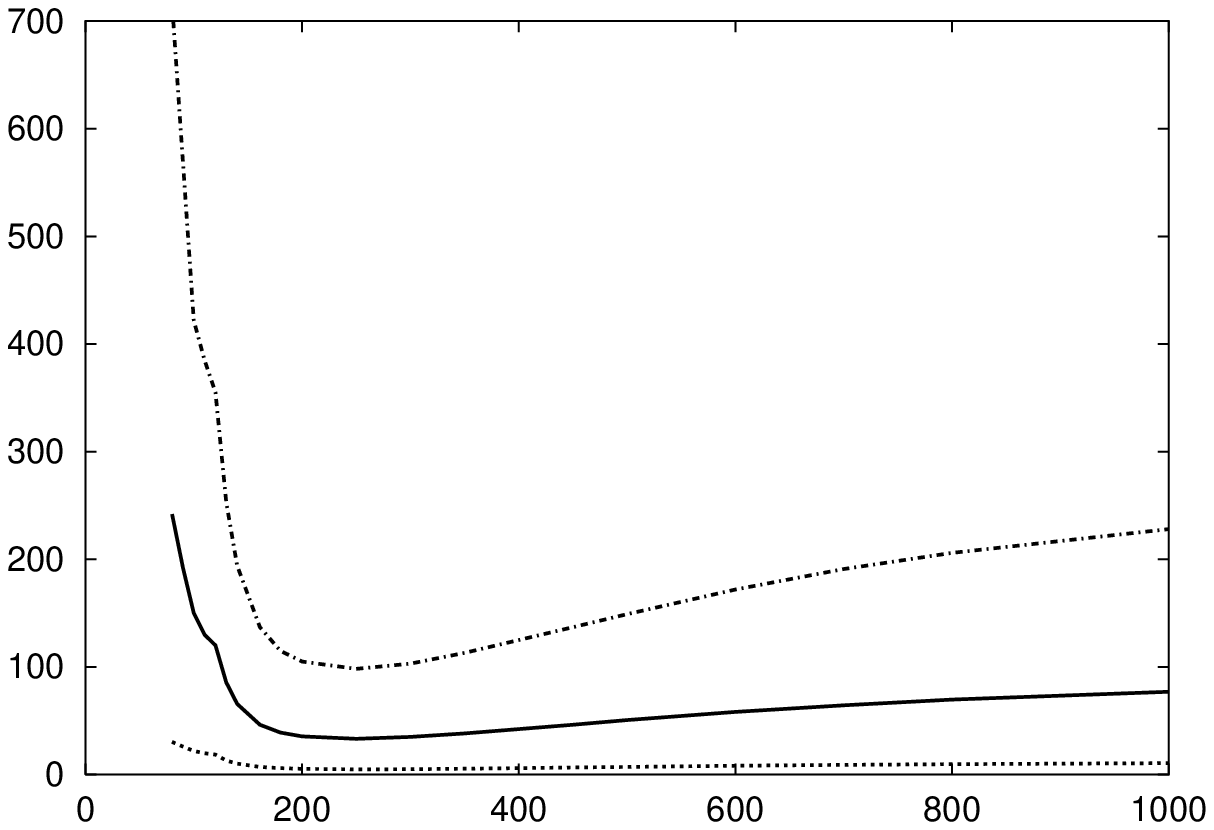}}
\put(290,-175){$ \scriptstyle m_{\tilde{\nu}_e}/GeV$}
\put(0,-18){$ \scriptstyle \sigma_{e^{-}} /fb$}
\put(0,-220)
{\parbox{11.5cm}{Fig.~2: Cross section $\sigma_{e^{-}}$ (see eq.~(\ref{sigma}))
    for $\sqrt{s}=500$~GeV in scenario~A for unpolarized beams (solid
    line), with  only $e^{-}$ beam polarized $P_{-}=+85\%$ (dotted
    line) and with both beams polarized 
    $P_{-}=-85\%$, $P_{+}=+60\%$ (dash-dotted line).}\label{sit500_A}}
\end{picture}

\newpage
\begin{picture}(10,8)
\put(4,0){\includegraphics{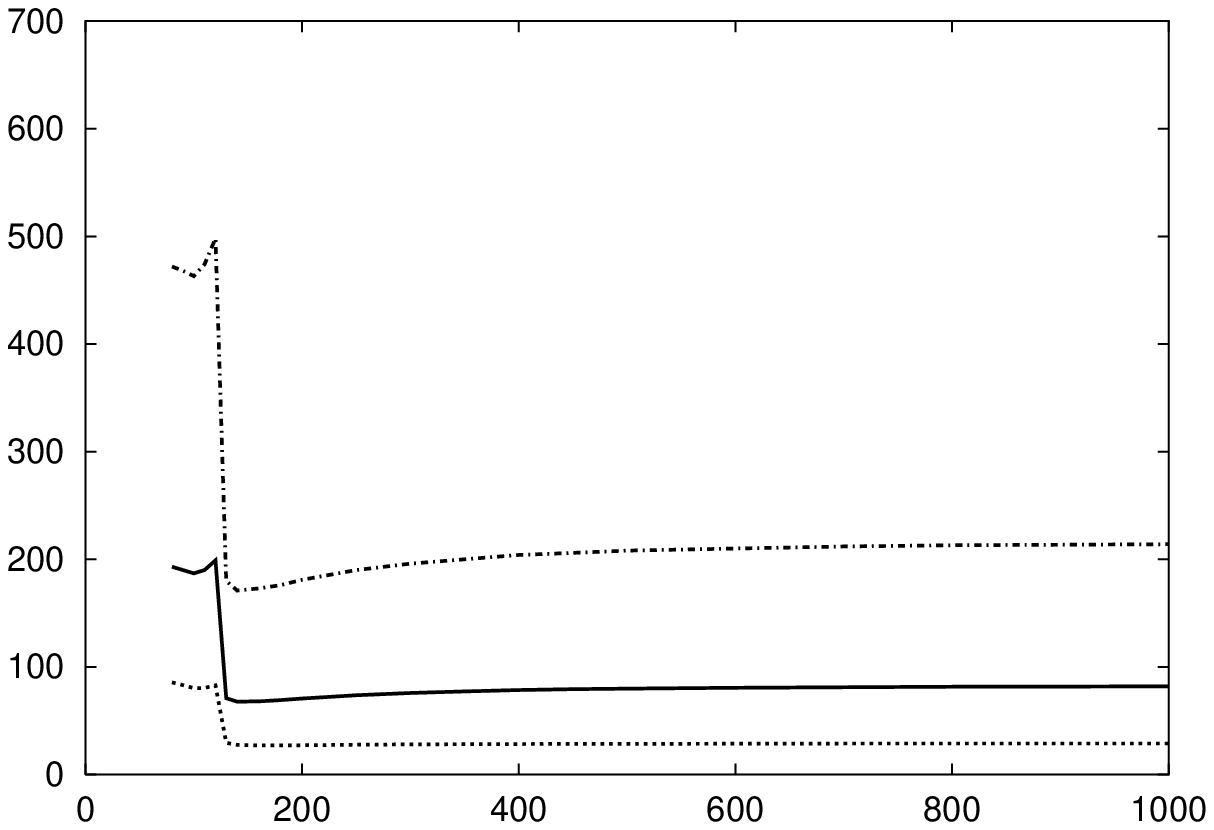}}
\put(290,-175){$ \scriptstyle m_{\tilde{\nu}_e}/GeV$}
\put(0,-18){$ \scriptstyle \sigma_{e^{-}} /fb$}
\put(0,-220)
{\parbox{11.5cm}{Fig.~3: Cross section $\sigma_{e^{-}}$ (see eq.~(\ref{sigma}))
    for $\sqrt{s}=270$~GeV in scenario~B for unpolarized beams (solid
    line), with  only $e^{-}$ beam polarized $P_{-}=+85\%$ (dotted
    line) and with both beams polarized 
    $P_{-}=-85\%$, $P_{+}=+60\%$ (dash-dotted line).}\label{sit270_B}}
\end{picture}

\vspace{8.5cm}
\begin{picture}(10,8)
\put(4,0){\includegraphics{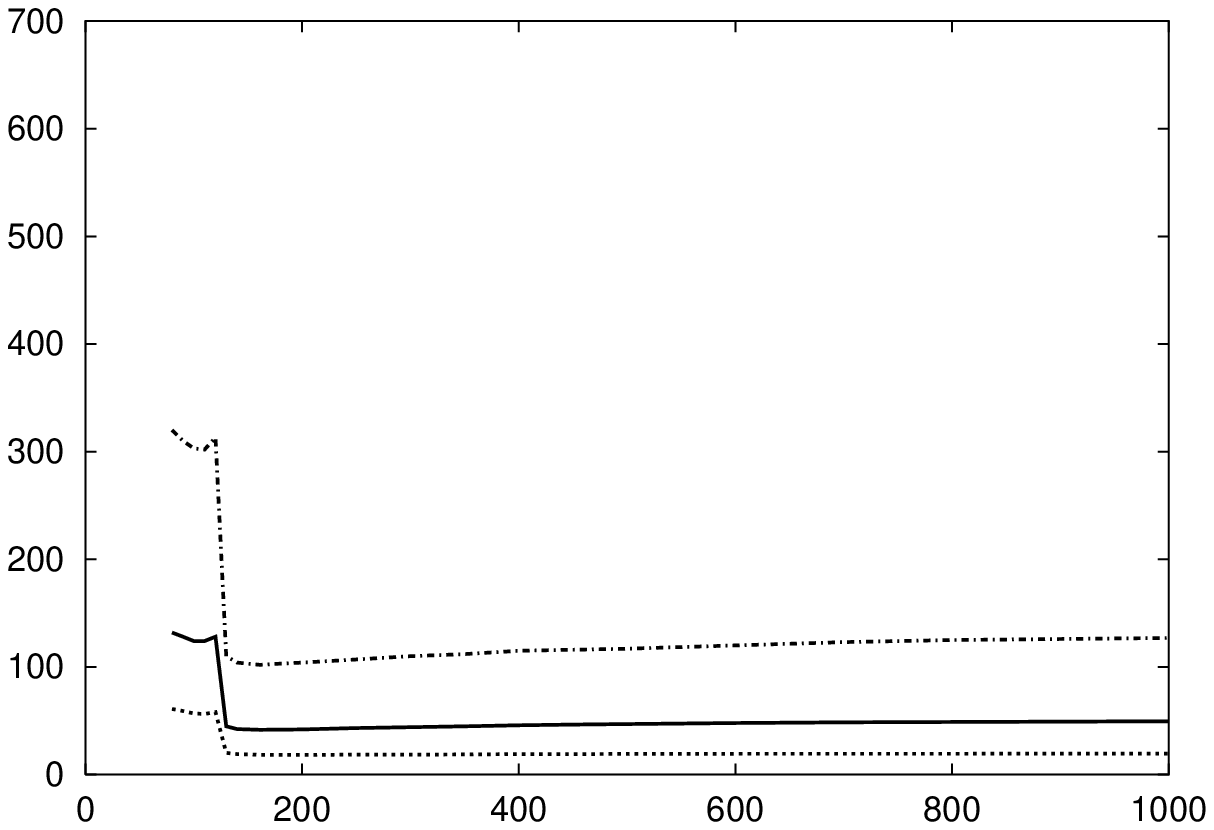}}
\put(290,-175){$ \scriptstyle m_{\tilde{\nu}_e}/GeV$}
\put(0,-18){$ \scriptstyle \sigma_{e^{-}} /fb$}
\put(0,-220)
{\parbox{11.5cm}{Fig.~4: Cross section $\sigma_{e^{-}}$ (see eq.~(\ref{sigma}))
    for $\sqrt{s}=500$~GeV in scenario~B for unpolarized beams (solid
    line), with  only $e^{-}$ beam polarized $P_{-}=+85\%$ (dotted
    line) and with both beams polarized 
    $P_{-}=-85\%$, $P_{+}=+60\%$ (dash-dotted line).}\label{sit500_B}}
\end{picture}
\newpage
\begin{picture}(10,8)
\put(-10,0){\includegraphics{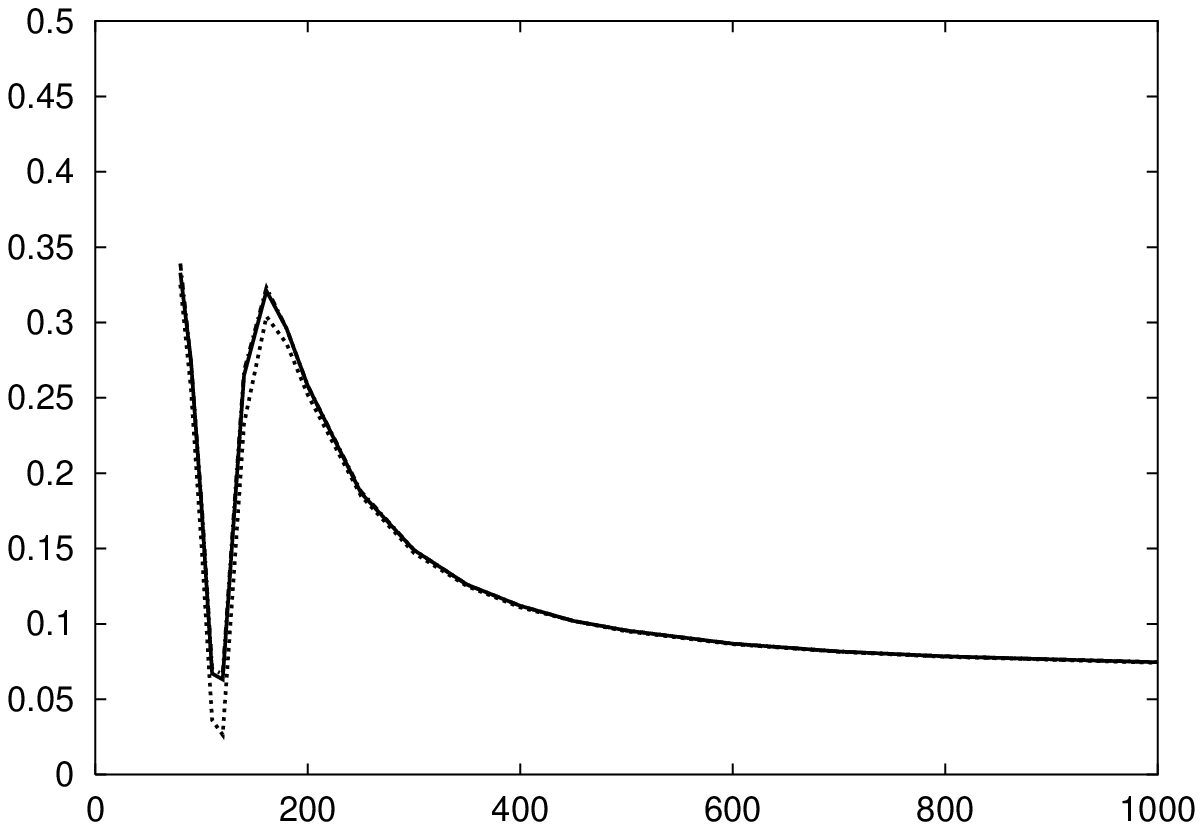}}
\put(280,-175){$ \scriptstyle m_{\tilde{\nu}_e}/GeV$}
\put(0,-18){$ \scriptstyle A_{FB}$}
\put(0,-220)
{\parbox{11.2cm}{Fig.~5: Electron forward--backward asymmetry $A_{FB}$ 
    (see eq.~(\ref{afb})) in the laboratory system
    for $\sqrt{s}=270$~GeV in scenario~A for unpolarized beams (solid
    line), with  only $e^{-}$ beam polarized $P_{-}=+85\%$ (dotted
    line) and with both beams polarized 
    $P_{-}=-85\%$, $P_{+}=+60\%$ (dash-dotted line).}\label{asy270_A}}
\end{picture}

\vspace{8.5cm}
\begin{picture}(10,8)
\put(-10,0){\includegraphics{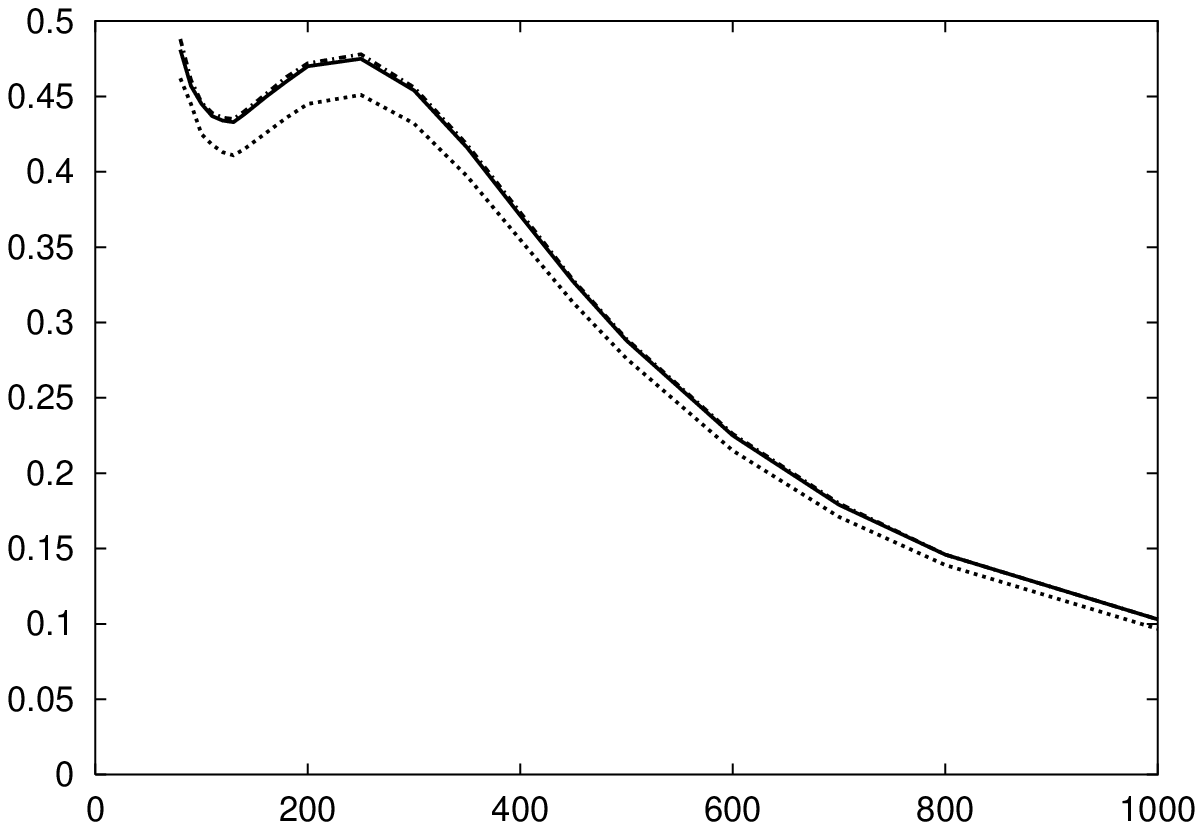}}
\put(280,-175){$ \scriptstyle m_{\tilde{\nu}_e}/GeV$}
\put(0,-18){$ \scriptstyle A_{FB}$}
\put(0,-220)
{\parbox{11.2cm}{Fig.~6: Electron forward--backward asymmetry $A_{FB}$
    (see eq.~(\ref{afb})) in the laboratory system
    for $\sqrt{s}=500$~GeV in scenario~A for unpolarized beams (solid
    line), with  only $e^{-}$ beam polarized $P_{-}=+85\%$ (dotted
    line) and with both beams polarized 
    $P_{-}=-85\%$, $P_{+}=+60\%$ (dash-dotted line).}\label{asy500_A}}
\end{picture}
\newpage

\begin{picture}(10,8)
\put(-10,0){\includegraphics{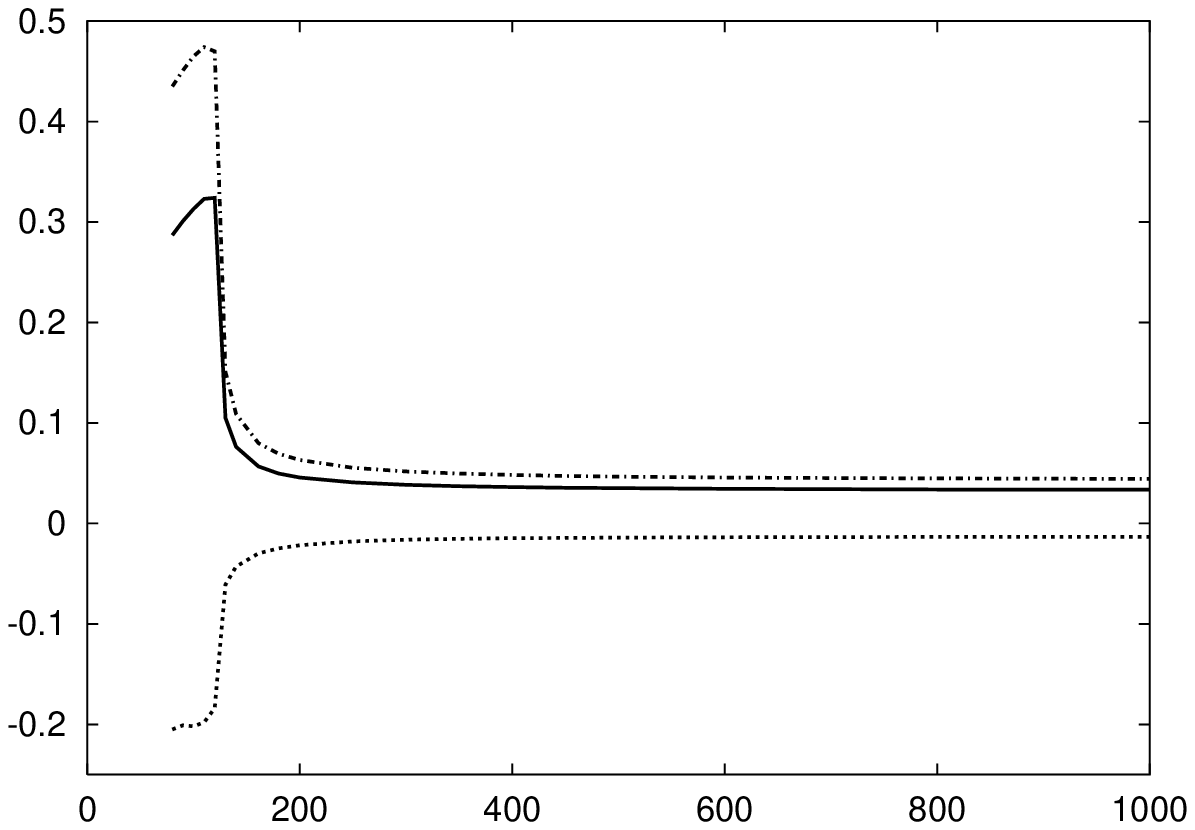}}
\put(280,-175){$ \scriptstyle m_{\tilde{\nu}_e}/GeV$}
\put(0,-18){$ \scriptstyle A_{FB}$}
\put(0,-220)
{\parbox{11.2cm}{Fig.~7: Lepton forward--backward asymmetry $A_{FB}$ 
    (see eq.~(\ref{afb})) in the laboratory system
    for $\sqrt{s}=270$~GeV in scenario~B for unpolarized beams (solid
    line), with  only $e^{-}$ beam polarized $P_{-}=+85\%$ (dotted
    line) and with both beams polarized 
    $P_{-}=-85\%$, $P_{+}=+60\%$ (dash-dotted line).}\label{asy270_B}}
\end{picture}

\vspace{8.5cm}
\begin{picture}(10,8)
\put(-10,0){\includegraphics{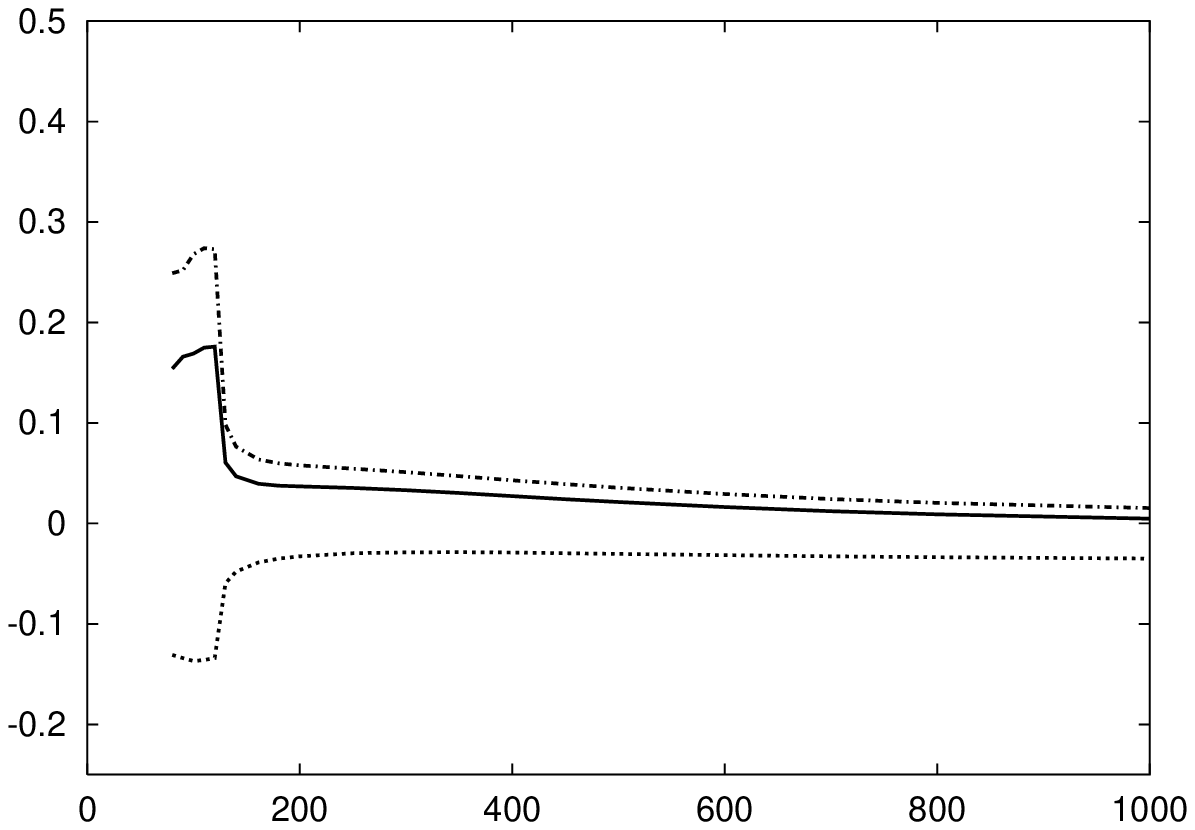}}
\put(280,-175){$ \scriptstyle m_{\tilde{\nu}_e}/GeV$}
\put(0,-18){$ \scriptstyle A_{FB}$}
\put(0,-220)
{\parbox{11.2cm}{ Fig.~8: Lepton forward--backward asymmetry $A_{FB}$ 
    (see eq.~(\ref{afb})) in the laboratory system  
    for $\sqrt{s}=500$~GeV in scenario~B for unpolarized beams (solid
    line), with  only $e^{-}$ beam polarized $P_{-}=+85\%$ (dotted
    line) and with both beams polarized 
    $P_{-}=-85\%$, $P_{+}=+60\%$ (dash-dotted line).}\label{asy500_B}}
\end{picture}
\end{document}